\documentclass{PoS}
\usepackage{wrapfig}
\title{Contribution title}
\title{LOFAR and HDF5: Toward a New Radio Data Standard}

\author{\speaker{Kenneth R. Anderson}
\thanks{A big thanks to all those who offered critique and commentary during this process.} \\
University of Amsterdam (UvA), Amsterdam, The Netherlands\\
 E-mail: \email{k.r.anderson@uva.nl}}

 \author{ Anastasia Alexov \\
 University of Amsterdam (UvA), Amsterdam, The Netherlands  \\
 E-mail: \email{a.alexov@uva.nl}}

\author{Lars B{\aa}hren \\
Radboud University of Nijmegen, Nijmegen, The Netherlands \\
E-mail: \email{lbaehren@gmail.com}}

\author{Jean-Mathias Grie{\ss}meier,\\
Netherlands Institute for Radio Astronomy (ASTRON), Dwingeloo, The Netherlands\\
E-mail: \email{griessmeier@astron.nl}}

 \author{Michael Wise \\
Netherlands Institute for Radio Astronomy (ASTRON), Dwingeloo, The Netherlands}

\author{Gerard Adriaan Renting \\
Netherlands Institute for Radio Astronomy (ASTRON), Dwingeloo, The Netherlands}
\abstract{For decades now, scientific data volumes have experienced
  relentless, exponential growth.  As a result, legacy astronomical
  data formats are straining under a burden not conceived when these
  formats were first introduced.  With future astronomical projects
  ensuring this trend, ASTRON and the LOFAR project are exploring the
  use of the Hierarchical Data Format, version 5 (HDF5), for LOFAR
  radio data encapsulation. Most of LOFAR's standard data products
  will be stored using the HDF5 format. In addition, HDF5
  analogues for traditional radio data structures such as visibility
  data and spectral image cubes are also being developed. The HDF5
  libraries allow for the construction of distributed,
  entirely unbounded files.  The nature of the HDF5 format further
  provides the ability to custom design a data encapsulation
  format, specifying hierarchies, content and attributes.
  The LOFAR project has designed several data formats that
  will accommodate and house all LOFAR data products, the primary
  styles and kinds of which are presented in this paper.  With proper
  development and support, it is hoped that these data formats will be
  adopted by other astronomical projects as they, too, attempt to
  grapple with a future filled with mountains of data.}

\FullConference{ISKAF2010 Science Meeting - ISKAF2010\\
		June 10-14, 2010\\
		Assen, the Netherlands}

\ShortTitle{LOFAR and HDF5}

\FullConference{ISKAF2010 Science Meeting - ISKAF2010\\
		June 10-14, 2010\\
		Assen, the Netherlands}

\begin{document}

\section{Introduction}
The promising advent of the LOFAR telescope's operational epoch holds forth both great scientifiic potential and challenges to current and legacy information technologies: volume and complexity of the data will continue to push the envelope of commonly used data protocols.

Recognizing that this envelope is already strained, the LOFAR project has embarked on an ambitious project to design and define a set of radio data standard formats that are capable of encapsulating the full spectrum of, not just LOFAR data products, but astronomical radio data in general.

It is with this ambition in mind that the LOFAR data formats group have been developing these format specifications and associated software infrastructure, an effort now ongoing for over two years.  It was determined that  HDF5 would be a robust, viable data framework that can handle the size, scope, diversity, distributed nature and parallel I/O processing requirements of LOFAR data. This work also has potential use beyond the radio community. New large scale optical telescopes such as the LSST are also investigating the viability of using HDF5. Furthermore, the 20 year history of HDF and its continuing use by NASA's earth orbiting/observing missions ensure broad, ongoing use and support.

 In addition to the format descriptions themselves, the LOFAR project is currently developing a set of software tools for creating and working with these formats. The Data Access Library (DAL) in C++, along with an associated Python interface (pyDAL), are designed to allow for the easy construction and manipulation of these data formats. There are also a number of tools already available to read and visualize HDF5 files, such as HDFView, ViSiT, PyTables, h5py and IDL.

\section{Data Files in the Modern Age}
With the invention and subsequent commercialization of the charge-coupled device (CCD), data volumes in the field of astronomy have grown exponentially since the early 1970s (Borne, 2009),  and, as Table 1 conveys, have now attained volumes that early file and file system protocols are finding difficult to handle.  For future radio astronomy projects, the looming plight is especially true.

\begin{table}[h]
  \centering
  \begin{tabular}{|cc|}
    \hline
    \sc Epoch & \sc Nominal file data volume\\ 
\hline \hline
 1970  &  2$^{10}$ bytes\\
 1980  &  2$^{20}$ bytes\\
 1990  &  2$^{30}$ bytes\\
 2000  &  2$^{40}$ bytes\\
\hline
\end{tabular}
\caption{Typical file sizes of scientific datasets have exponentially increased
since the 1970's.}
\end{table} 

Indeed, the path of diminishing utility of legacy protocols is clearly delineated.  This looming predicament is especially germane to the SKA pathfinder LOFAR project, wherein certain operational modes will be capable of generating datasets comprising hundreds of gigabytes to tens of terabytes.  Therefore, the LOFAR project has been driven to consider viable alternatives to ``standard'' astronomical data formats, such as FITS.  Though FITS development has sought to add capability to the protocol as needed (ESO-FWG, 2008), such ex post facto approaches should not be expected to remain viable in the future.

\begin{table}[hbt]
  \centering
  \begin{tabular}{|ccccc|}
    \hline
    \sc Exposure & \sc Number of & \sc Number of & \sc File Size  & \sc File Size  \\
    \sc Time     & \sc Subbands  & \sc Stations  & \sc Known Mode & \sc Search Mode \\
    \hline \hline
     1 min & 248 & 5 & 11.2GB & 56GB \\
     1 min & 248 & 20 & 11.2GB & 244GB \\
\hline
     10 min & 248 & 5  & 112GB & 560GB \\
     10 min & 248 & 10 & 112GB & 1.1TB \\
     10 min & 248 & 20 & 112GB & 2.2TB \\
     10 min & 248 & 30 & 112GB & 3.3TB \\
\hline
     20 min & 248 & 5 & 224GB & 1.1TB \\
\hline
     30 min & 248 & 5 & 336GB & 1.7TB \\
\hline
     1 hr  & 248 & 5  & 672GB & 3.4TB \\
     1 hr  & 248 & 10 & 672GB & 6.7TB \\
     1 hr  & 248 & 20 & 672GB & 13.4TB\\
     1 hr  & 248 & 30 & 672GB & 26.8TB \\
\hline
     2 hr  & 248 & 5 & 1.3TB & 6.7TB \\
\hline
     12 hr & 248 & 5 & 8.0TB & 40.3TB \\
     12 hr & 248 & 15 & 24.0TB & 120.1TB \\
    \hline
 \end{tabular}
  \caption{Sample, LOFAR Beam-formed dataset sizes}
  \label{tab:data size}
\end{table}

Recognizing that legacy file technologies will eventually fail under the growing
demands of newer and larger scientific datasets, while simultaneously noting that if observational modes of SKA pathfinder projects like LOFAR suggest anything, it is that legacy technologies would have to be abandoned, not just because of the volumes of data involved, but also because of the complex nature of the data per se.

By way of LOFAR example, Table 2 presents expected data
volumes for configurations of Beam-Formed data observational mode.  A
once useful unit, the megabyte has been left in the wake of next
generation science. 

\section{LOFAR Data Formats in HDF5}
\subsection{LOFAR data and the use HDF5 (Hierarchical Data Format 5)}

Datasets produced by LOFAR observations will vary tremendously in
size.  Images, Beam-formed data, Transient Buffer board (TBB)
time-series data are expected to produce large files, with the
beam-formed and TBB potentially forming files of several tens of terabytes.  This, combined with the complex nature of the data from certain modes of observation, led the LOFAR project to examine and consider the Hierarchical Data Format (version 5) as a robust and viable solution to the volume-complexity problem:\\\\-- HDF5 presents a robust data model, featuring distributed files.\\
-- HDF5 data model: ``a file system for your data.'' which operates through POSIX-style file navigation, e.g.,  path names to objects.\\
\verb|     OBSERVATION/Image001/Data|\\
-- Versatile data model accommodates complex data objects and associated metadata.\\
-- User-defined file design.\\
-- Portable, unbounded file format.\\
 -- Multi-platform software library for single to massively parallel systems.\\

The following schematic diagrams present two  LOFAR file designs using the HDF5 framework, while readers should bear in mind that all LOFAR data products have been designed with maximal parallel form.

\begin{figure}[htbp]
  \centering
  \includegraphics[scale=0.52]{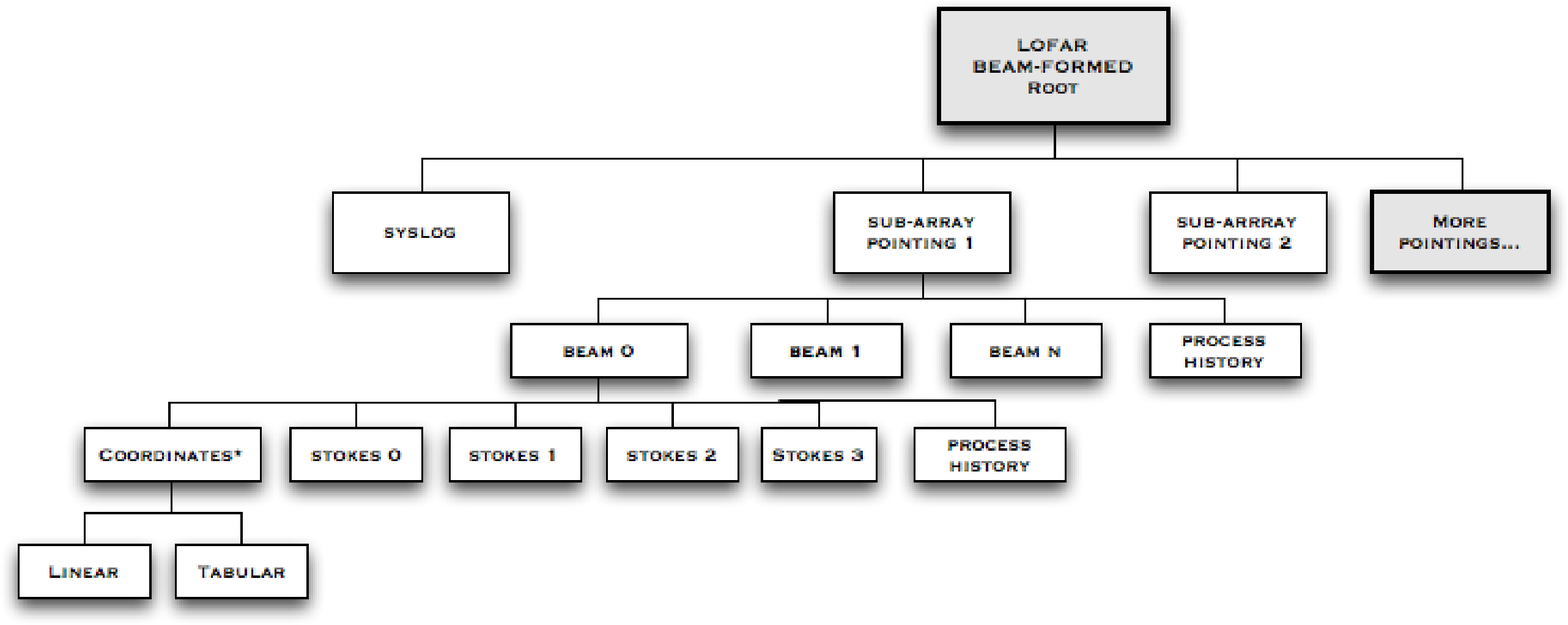}
  \caption{LOFAR Beam Formed High Level Data Structure; tables/arrays are not shown, but are implied at the Stokes level. (Alexov et al, 2010)}
  \label{fig:bfhighlev}
\end{figure}

\begin{figure}[htbp]
  \centering
  \includegraphics[scale=0.52]{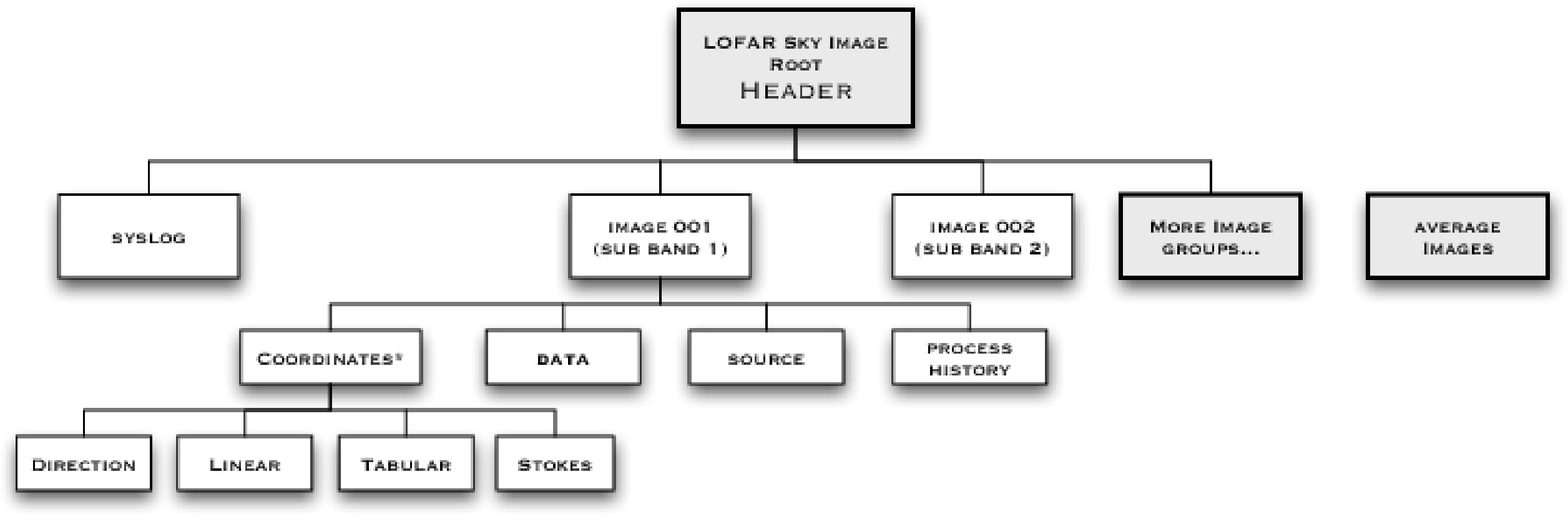}
  \caption{LOFAR Radio Sky Image Cubes, High Level Data Structure; tables/arrays are not shown, but are implied at the Data group level. (Anderson et al, 2010)}
  \label{fig:bfhighlev}
\end{figure}

\section{LOFAR Data Format Specifications}
HDF5 provides a framework allowing users to essentially design their
own files to  appropriately encapsulate and characterize a variety of data.

For ongoing years now, the LOFAR project has been
engaged in  developing and designing a complete set of specifications
for all LOFAR  observational data.  This has necessarily required
differing file designs  for differing data, with a certain structural
parallelism maintained  across all file designs.  These \textbf{Interface Control Documents (ICD)} (shown with their document indentifiers)  provide detailed descriptions of all  expected LOFAR Data Products.\footnote{Interface control documents LOFAR-USG-ICD-002, and LOFAR-USG-ICD-005 provide supplemental specifications for LOFAR naming conventions, and therefore are not included in the listing.}
\begin{table}[h]
  \centering
  \begin{tabular}{|ll|}
    \hline
    \sc LOFAR Data File type& \sc LOFAR Document ID\\ 
\hline \hline
\verb|TBB Time Series Data|  &\verb|LOFAR-USG-ICD-001|\\
\verb|Beam-Formed Data|          &\verb|LOFAR-USG-ICD-003|\\
\verb|Radio Sky Image Cubes|&\verb|LOFAR-USG-ICD-004|\\
\verb|Dynamic Spectrum Data|&\verb|LOFAR-USG-ICD-006|\\
\verb|Visibility Data|            &\verb|LOFAR-USG-ICD-007|\\
\verb|RM Synthesis Cubes|      &\verb|LOFAR-USG-ICD-008|\\
\hline
\end{tabular}
\caption{LOFAR file types and the corresponding Interface Control Documents}
\end{table} 
\\All LOFAR ICD documents are publicly available and can be downloaded from\\\\ \verb|http://usg.lofar.org/wiki/doku.php?id=documents:lofar_data_products|

\section{Toolsets, Libraries, Packages}
Since inception in 1988,  the body of libraries and tools available to work with HDF5 files has grown substantially.  Continued use, maintainance and development is assured with the adoption of HDF by Boeing, NASA -- for the Earth Observing System (EOS) -- while the National Oceanographic and Atmoshperic Administration adopted HDF for the National Polar-orbiting Operational Environmental Satellite System, NPOESS.  

In addition to the HDF5 library itself,  these packages and interfaces are also available:
\begin{table}[h]
  \centering
  \begin{tabular}{|ll|}
    \hline
    \sc Package& \sc Location\\ 
\hline \hline
 IDL &\verb|http://www.ittvis.com/|\\
ViSit&\verb|https://wci.llnl.gov/|\\
HDFView  &\verb|http://www.hdfgroup.org/|\\
DAL&\verb|http://usg.lofar.org/|\\
pyDAL&\verb|http://usg.lofar.org/|\\
h5py&\verb|http://h5py.alfven.org/|\\
PyTables&\verb|http://www.pytables.org/|\\
\hline
\end{tabular}
\caption{Some available HDF5 packages}
\end{table} 

The LOFAR project is developing the C++ \textbf{Data Access Library (DAL)} and an associated python wrapper, pyDAL, which  will provide full scope constructors for creating and accessing LOFAR data products.  Further tools and packages are catalogued at the HDF Group website, \verb|http://www.hdfgroup.org/tools5desc.html|

\section{Summary and Future Considerations}

In order that adoption of HDF5 in astronomy prove useful in the real
world, the LOFAR  project is committing resources to help develop the
next generation  of astronomical tools for, not only LOFAR data, but other SKA pathfinder projects and, more broadly, astronomical radio data in general.  Though our immediate goal is, of course, to meet LOFAR's many and varied scientific requirements, we hope these HDF5-based formats will be more broadly useful to our colleagues in the radio community for their data products as well.

The large effort is the development of the Data Access Library (DAL),
which will ultimately provide interfaces through FITS, the
Casa/AIPS++ Measurement Set  and LOFAR HDF5 formats.  A python interface to the
DAL, pyDAL, is pending development of the DAL.   All LOFAR products will be
accessible through DAL and pyDAL tools. 

Future work involves developing an interface for DS9 and HDF5 LOFAR
data files --- this will allow users to open and examine LOFAR data
with the de facto standard  astronomical image viewer.

Ultimately we would like to see these formats grow into a true set of standards for radio data that can meet the demands of the next generation of radio observatories. Such standards are something sorely lacking in the radio community at present and something we will certainly need as we move into the SKA era.

The large effort by LOFAR to design an HDF5 radio data standard is
driven in great part by consideration that there are no
effective standards for astronomical radio data.  And, as indicated earlier,
the expected data volumes produced by LOFAR will, in many cases,
swamp currently employed file technologies.  It is this reality that has led LOFAR
on this work, and it is hoped that other institutes and telescope projects will join this effort toward building a radio data standard, one essential to a cooperative future for
radio astronomy.

\end{document}